\documentclass[pre,twocolumn,superscriptaddress,amsmath,amssymb,showpacs,noshowkeys,a4paper]{revtex4}
\usepackage{amsmath}
\usepackage{graphicx}
\usepackage{dcolumn}
\usepackage{bm}
\usepackage[usenames]{color}
\usepackage{epstopdf}
\definecolor{mygray}{gray}{0.5}
\newcommand\vR{\mathbf{R}}

\newcommand\vradius{\mathbf{r}}

\newcommand\vv{\mathbf{v}}
\newcommand\vT{\mathbf{T}}

\newcommand\vN{\mathbf{N}}

\newcommand\etal{{\it et al.}}

\newcommand\ex{\mathbf{e}_x}
\newcommand\ey{\mathbf{e}_y}

\newcommand\bnabla{\bm{\nabla}}

\newcommand{\afflangevin}{\address{Institut Langevin, ESPCI Paristech, CNRS - UMR 7587, Universit\'e Pierre and Marie Curie,  1 rue Jussieu, 75005, Paris, France, EU}}
\newcommand{\afflangevinbis}{\address{Institut Langevin, ESPCI Paristech, CNRS - UMR 7587, PSL Research University,   1 rue Jussieu, 75005, Paris, France, EU}}

\newcommand{\affMSC}{\address{Laboratoire Mati\`{e}re et Syst\`{e}mes Complexes, Universit\'e Paris Diderot, Sorbonne Paris Cit\'{e}, CNRS - UMR 7057, 10 Rue A. Domon and L. Duquet, 75013 Paris, France, EU}}

\begin{document}
\title{Build-up of macroscopic eigenstates in a memory-based constrained system}
\author{M. Labousse}
\afflangevin
\affMSC
\author{S. Perrard}
\affMSC
\author{Y. Couder}
\affMSC
\author{E. Fort}
\email{emmanuel.fort@espci.fr}
\afflangevinbis
\begin{abstract}
A bouncing drop and its associated accompanying wave forms a walker. Based on previous works, we show in this article that it is possible to formulate a simple theoretical framework for the walker dynamics. It relies on a time scale decomposition corresponding to the effects successively generated when the memory effects increase. While the short time scale effect is simply responsible for the walker's propulsion, the intermediate scale generates spontaneously pivotal structures endowed with angular momentum. At an even larger memory scale, if the walker is spatially confined, the pivots become the building blocks of a self-organization into a global structure. This new theoretical framework is applied in the presence of an external harmonic potential, and reveals the underlying mechanisms leading to the emergence of the macroscopic spatial organization  reported by Perrard \textit{et al.}~\cite{Perrard_natureC_2014} (2014 {\it Nature Commun.} {\bf 5} 3219).
\end{abstract}

\pacs{47 55.D- Drops,  05 45.-a, Nonlinear dynamics and chaos}
\maketitle

\section{Introduction}
Complex systems often require building several hierarching levels of description. This raises fundamental questions in high-dimensional systems about the relevant amount of information needed to characterize them. \cite{Farmer_1982,Takens_1981,Wolf_1985}. Here, we investigate the case of a strongly nonlinear dynamics in which its apparent complexity can be tremendously reduced, giving way to the surprising quantification of its observables. 
\\

An experimental situation has revealed that a \textit{walker}~\cite{Walker_Nature}, this macroscopic association of a bouncing drop and its accompanying wave, exhibits a dynamics sufficiently complex to reproduce at an unusual scale, a quantum-like effect such as diffraction through submarines slits \cite{Couder_Diffraction}, tunneling \cite{Eddi_Tunnel}, Zeeman-like splitting \cite{Eddi_PRL_2012},wave-like statistics in cavities \cite{Harris_PRE_2013} or else Landau level analogue \cite{Fort_PNAS,Oza_JFM_2_2013,Harris_JFM_2014}. Perrard \etal  \cite{Perrard_natureC_2014} investigated the case where the system, a droplet set on a vertically-vibrated bath and self-propelled by its accompanying wave, is subjected to a two-dimensional (2D) harmonic central potential of natural frequency $\omega/2\pi$. They singled out elementary paths on which the full dynamics can be decomposed. In particular, they showed that the phase space of the dynamics can be projected on a state diagram $(n,m)$. The integer $n$ is associated with the time-average orbit extension and $m$ the quantized  mean angular momentum, satisfying $m \in \lbrace -n, -n+2, \ldots, n-2,n\rbrace$ reminiscent of the quantum selection rules. In the current paper, we investigate the mechanism responsible of the emergence of theses macroscopic eigenstates. \\

As in the previous reported experiments, we consider a millimetric drop of silicon oil bouncing on a bath oscillating vertically at a frequency $f_0=80$ Hz with an acceleration $\gamma=\gamma_m \cos( 2 \pi f_0 t)$. The coalescence is prevented by the permanent presence of a thin film of air between the drop and the bath \cite{JFM_Suzie,Terwagne_Phys_2008,Molacek_JFM_1_2013, Molacek_POF_2012}. In addition, above a critical Faraday acceleration threshold $\gamma_F$, the surface is unstable at a frequency $f_0/2$ and standing waves appears spontaneously as initially observed by Faraday \cite{FaradayOriginal} and further studied by \cite{Benjamin_Ursell,Douady_JFM,EPLDouadyFauve}  (see the review \cite{Miles_1990} and references therein). We restrain our study to the case with period doubling at an acceleration amplitude $\gamma_m$ slightly below the Faraday acceleration threshold, typically $\gamma_F/(\gamma_F-\gamma_m) \sim 0.98$. In this situation, the drop excites parametrically the surface wave, each impact generating a Bessel-like mode centered at the impact point \cite{Eddi_JFM_2011,Molacek_JFM_2_2013}. The surface height $h$ derives from the linear superposition of the elementary contributions of the previous impacts. Each contribution relaxes to the equilibrium over a typical time decay, henceforth the memory time $\tau=M T_F$. This memory parameter $M$ indicating the relevant number of secondary sources contributing to the surface field. $M\sim \gamma_F/(\gamma_F-\gamma_m)$ is tuned by changing the driving acceleration amplitude. Impacts that
occurred more than a few $M$ Faraday periods in the past are negligible. Above a critical memory parameter $M_c\sim 3$, the vertical bouncing state is unstable to a horizontal perturbation. The drop lands on an inclined surface and acquires an increment of horizontal momentum proportional to the local slope \cite{Fort_PNAS,Molacek_JFM_2_2013,Oza_JFM_1_2013}, which initiates a horizontal motion at typical speed $\sim 10$ mm/s.\\

\begin{figure*}
    \centering
\includegraphics[width=\textwidth]{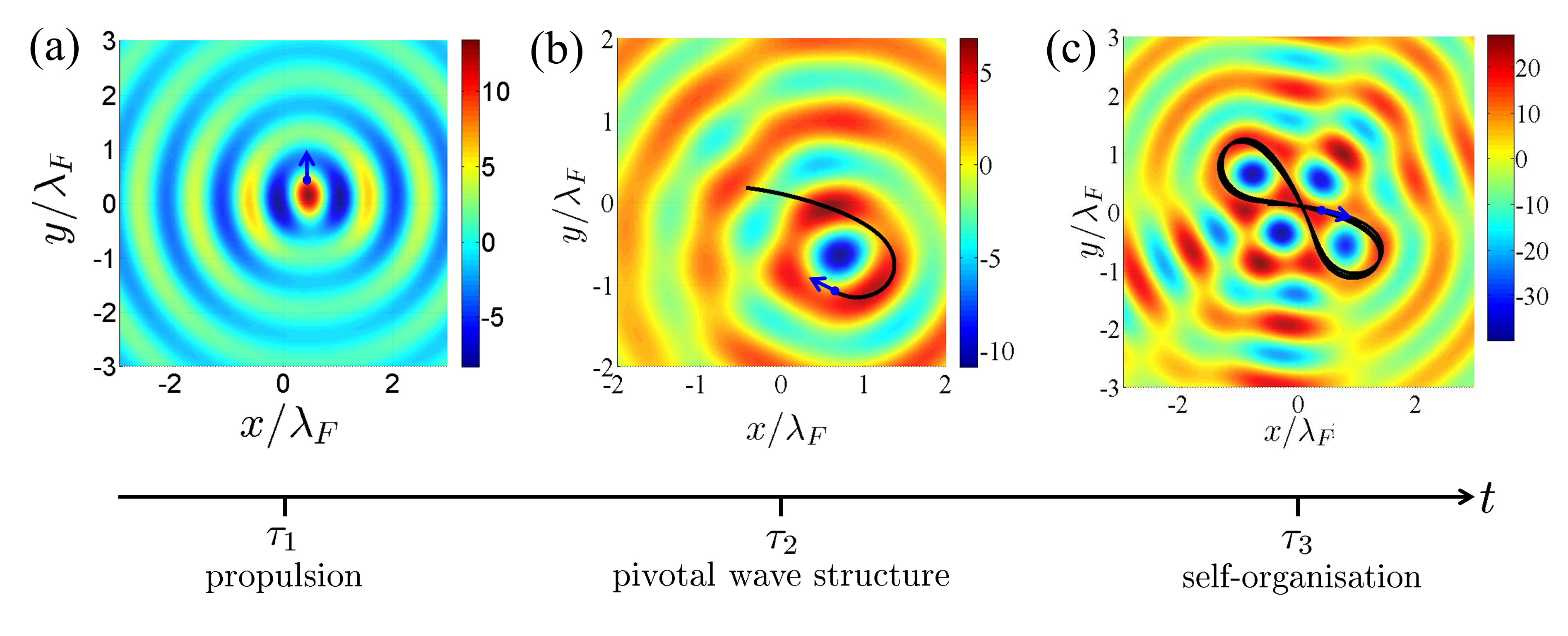}
\caption{Typical example of a walker dynamics in a harmonic potential (here $(\omega/2\pi=1.32 Hz)$ in natural unit or $(\omega/2\pi= 0.0330)$ in Faraday period time unit) and the corresponding wave field as the memory increases. In this specific case the time average speed is $7.9$ mm/s. The wave fields are reconstructed from the simulated paths (Fort numerical model~\cite{Perrard_natureC_2014}). (a) Local structure: Memory $M=$9, the wave field propels the drop. (b) Semi local structure: $M=$19, some semi local pivotal structures arise. The drop is propelled and guided by a pivotal surface wave field. (c) Global structure: $M=$150, the drop is now propelled and guided in a fully coherent structure, corresponding to a well-defined organization of pivotal surface wave field.}
\label{Multiscale}
\end{figure*}
Figure \ref{Multiscale} represents typical examples of the wave field and the associated horizontal walker dynamics as the memory increases. For short memory parameter $M$ (see Figure \ref{Multiscale}(a)), the surface field mainly propels the droplet forward at a constant speed. This regime defines the short time scale $\tau_1 \sim T_F$ and the eponymous dynamics. For intermediate memory parameter $M$ (see Figure \ref{Multiscale}(b)), an increasing number of secondary sources can interfere and semi-local surface field structures can arise. The walker can turn around these structures and reinforces them. We unambiguously call these characteristic wave structures, \textit{pivotal wave structures}. This phenomenon acts on a typical time scale $\tau_2\gg \tau_1$ and defines the intermediate time scale dynamics.  As the memory is much larger (see Figure \ref{Multiscale}(c)), several reminiscent pivotal fields can coexist and a global coherent structure emerges from an apparently complex path. It corresponds to a well-defined organization of these pivotal wave structures. This organization acts on a third time scale $\tau_3 \gg \tau_2\gg\tau_1 $ and defines the long time scale dynamics. It indicates a clear separation of time scale, associating with space scale organization of well-defined surface field structures. At long memory, all these time scales interlock and the effect of each of them can be revealed as the memory parameter increases. The main goal of this paper is to present a theoretical framework in the presence of an attractive potential in the light of this spatio-temporal separation of scales. As it is surprising that this macroscopic wave particle association, when immersed into a harmonic potential, gives rise to a set of attractors with classical selection rule reminiscent of its quantum counterpart \cite{Perrard_natureC_2014}, we propose to apply this theoretical framework to explain how such classical attractors emerge.\\

In a first part, we recall the symmetry properties of the surface field and their consequence on its further time-scale decomposition. In a second part, we formulate the short time dynamics and express it close to the constraint of small speed fluctuations. In a third part, we zoom out a first time and we add the semi-local wave structure emerging from the pivotal field. The dynamics is described in the Frenet-adapted wave basis and show that this basis is actually adapted to the translational invariance property of the surface field. We describe how the system evolves with preferred radii of curvature. These pivotal structures constitute the basic units of the dynamics. In a last part, we zoom out once again and see how these pivots interact and organize with each other, which defines the long time dynamics.  We show that this self organization emerges from a compromise between two \textit{a priori} incompatible symmetries. The location of the translational-invariant pivotal fields have to account for the rotational-invariant central force. 

\section{The space-time separation of the field \label{intro}} 
 The theoretical description of the walker dynamics has already been introduced in previous works \cite{Fort_PNAS,Walker_Nature,Eddi_JFM_2011,Molacek_JFM_1_2013,Molacek_JFM_2_2013,Oza_JFM_1_2013}. The idea of this paper is to use the time scale separation of the dynamics to reduce the complexity of the existing theoretical frameworks. As the drop bounces synchronously with the bath every Faraday period, and the horizontal distance between two impacts is much smaller than the Faraday wavelength, the horizontal dynamics can be approximated by a continuous description \cite{Oza_JFM_1_2013}
\begin{equation}
\frac{d\vv}{dt}=-\gamma \vv -C \left[\bnabla h\right]_{\vradius(t)}-\left[\frac{\bnabla E_p}{m}\right]_{\vradius(t)}
\end{equation}
with $\vradius$ the horizontal drop position, $\vv$, its horizontal speed as sketched in Figure \ref{schema}(a). The length scales and time are normalized by $\lambda_F/2\pi$ ($\lambda_F=4.75$ mm) and the Faraday period $T_F$ ($T_F=25$ ms). $E_p$ denotes an external attractive potential that will be further specified. The interaction with the surface consists in an apparent friction term  -$\gamma \vv $ and a coupling with the local slope of the surface field $-C \left[\bnabla h\right]_{\vradius(t)}$, the gradient being taken at the drop position at a given time. In principle, the coefficients $\gamma$ and $C$ can be deduced from the benchmarked Fort's numerical model~\cite{Perrard_natureC_2014,Couder_Diffraction,Fort_PNAS} or from the hydrodynamic model of Mol\'a\v{c}ek \textit{et al.}~\cite{Molacek_JFM_1_2013,Molacek_JFM_2_2013}. They are used here as free parameters and further chosen to match the experiments. Evaluating this field at the instantaneous particle position $\vradius(t)$ yields \cite{Fort_PNAS,Oza_JFM_1_2013}. 
\begin{equation}
\displaystyle h\left(\vradius,t \right) =\int_{- \infty }^{t}\frac{dT}{T_F} J_0\left(\Vert \vradius(t)-\vradius(T) \Vert \right)e^{-(t-T)/M}
\label{integralechamp}
\end{equation}
with $J_0$ the first order Bessel function, being centered at the point of a past impact $\vradius(T)$ and felt at $\vradius(t)$, the current position of the drop. The origin of different time scales arises intrinsically from the surface term. It is typically the time interval required to generate different hierarchies of coherent wave structures. In paragraph \ref{intro1}, we first recall symmetry properties of the surface field $h$ and we use them in paragraph \ref{intro2} to separate the different time scales of the dynamics.
\subsection{Symmetry properties of the surface field \label{intro1}}
\begin{figure*}
    \centering
\includegraphics[width=\columnwidth]{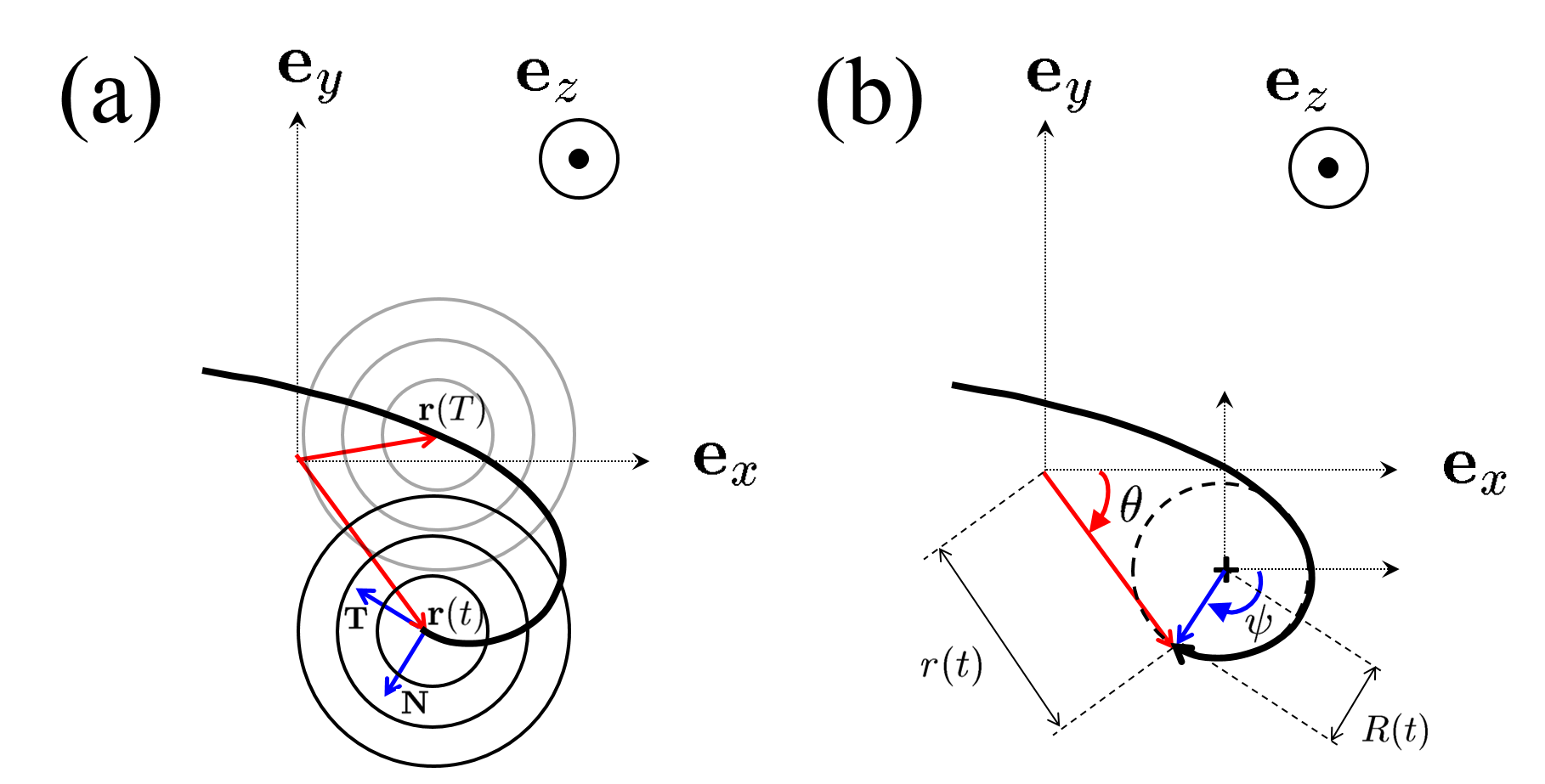}
\caption{(a) Schematics of the dynamics. The origin of referential $(\ex,\ey)$ corresponds to the minimum of energy of the attractive potential. The motion defined the tangential and the normal direction so that $(\vT,\vN)$ forms a direct basis. The standing wave-field at a given instant $t$ interferes with the remaining standing wave generated in the past (For all past sources so that $T<t$) . (b) We define two bases: the polar basis $(r,\theta)$ of origin prescribed by the minimum of energy of the attractive potential and the basis $(R,\psi)$ of origin at the instantaneous center of curvature. $R$ denotes the algebraic radius of curvature}
\label{schema}
\end{figure*}
The dynamics relies on three difference time scales, the propulsion, the semi-local structure and the global organization. Zooming in on these different dynamical aspects requires developing different tools, particularly in order to describe the surface field.\\

Being a solution of the two dimensional Helmholtz equation, the local value of surface field can be decomposed into polar Bessel eigenfunction $\lbrace f_n(\tilde{r},\tilde{\theta})\rbrace_{n\in \mathbb{Z}} =\lbrace J_n(\tilde{r})e^{in\tilde{\theta}})\rbrace _{n\in \mathbb{Z}}$, with a free choice in the center of decomposition. In this sense, the wave field revealed translational-invariant properties. The surface field can be indifferently decomposed onto a Frenet-adapted wave basis $\lbrace f_n^F(t)\rbrace_{n\in \mathbb{Z}} =\lbrace J_n(R(t))e^{in\psi(t)}\rbrace _{n\in \mathbb{Z}}$
  
\begin{equation}
\left\{
\begin{array}{ll} 
    \displaystyle h\left(\vradius(t),t \right) =\sum\limits_{n\in \mathbb{Z}} h_n^{F} f_n^F(t) \\
    \displaystyle h_n^{F}(t)=\langle h \vert f_{n}^F \rangle=\int\limits_{- \infty }^{t} \frac{dT}{T_F} J_n\left(R(T) \right) e^{-in\psi(T)} e^{-(t-T)/M}
        \end{array}
\right.
\label{decompositionchampsemilocal}
\end{equation}
as well as onto a basis of center adapted to the external potential $\left\lbrace f_n^{\mathrm{ext}}(t)\right\rbrace_{n\in \mathbb{Z}} =\left\lbrace J_n(r(t))e^{in\theta(t)}\right\rbrace _{n\in \mathbb{Z}}$
 \begin{equation}
\left\{
\begin{array}{ll} 
    \displaystyle h\left(\vradius,t \right) =\sum\limits_{n\in \mathbb{Z}} h_n^{\mathrm{ext}} f_n^{\mathrm{ext}}(t)\\
    \displaystyle h_n^{\mathrm{ext}}(t)=\langle h \vert f_{n}^{\mathrm{ext}} \rangle=\int\limits_{- \infty }^{t} \frac{dT}{T_F} J_n\left(r(T) \right) e^{-in\theta(T)} e^{-(t-T)/M}
\end{array}
\right.
\label{decompositionchampglobal}
\end{equation}
The different coordinates are indicated in Figure \ref{schema}(b) and correspond to two different polar bases. In equation \ref{decompositionchampsemilocal}, the Frenet basis $R$ denotes the instantaneous radius of curvature while $\psi$ indicates the local polar angle having its origin at the instantaneous center of curvature. In the central polar basis decomposition (equation  \ref{decompositionchampglobal}), $(r,\theta)$ indicates the usual polar coordinates, the origin coinciding with the center of the external potential. Let us briefly mention that in both equations \ref{decompositionchampsemilocal} and \ref{decompositionchampglobal} no imaginary part is effectively added and the whole expression remains real.\\

Equations \ref{decompositionchampsemilocal} and \ref{decompositionchampglobal} are mathematically equivalent and simply correspond to two different viewpoints. The decomposition in the Frenet basis (equation \ref{decompositionchampsemilocal}) corresponds to a projection of the local value of the wave field into eigenmodes. This decomposition will be used in Sec.\ref{intermediate} to express the intermediate time scale dynamics. The decomposition onto a central basis reflects the symmetry of the harmonic potential. We will study in paragraph \ref{Long} how the pivotal wave structure builds a long term self-organization.

\subsection{Time scale decomposition of the surface field\label{intro2}}
\begin{figure*}
    \centering
\includegraphics[width=\columnwidth]{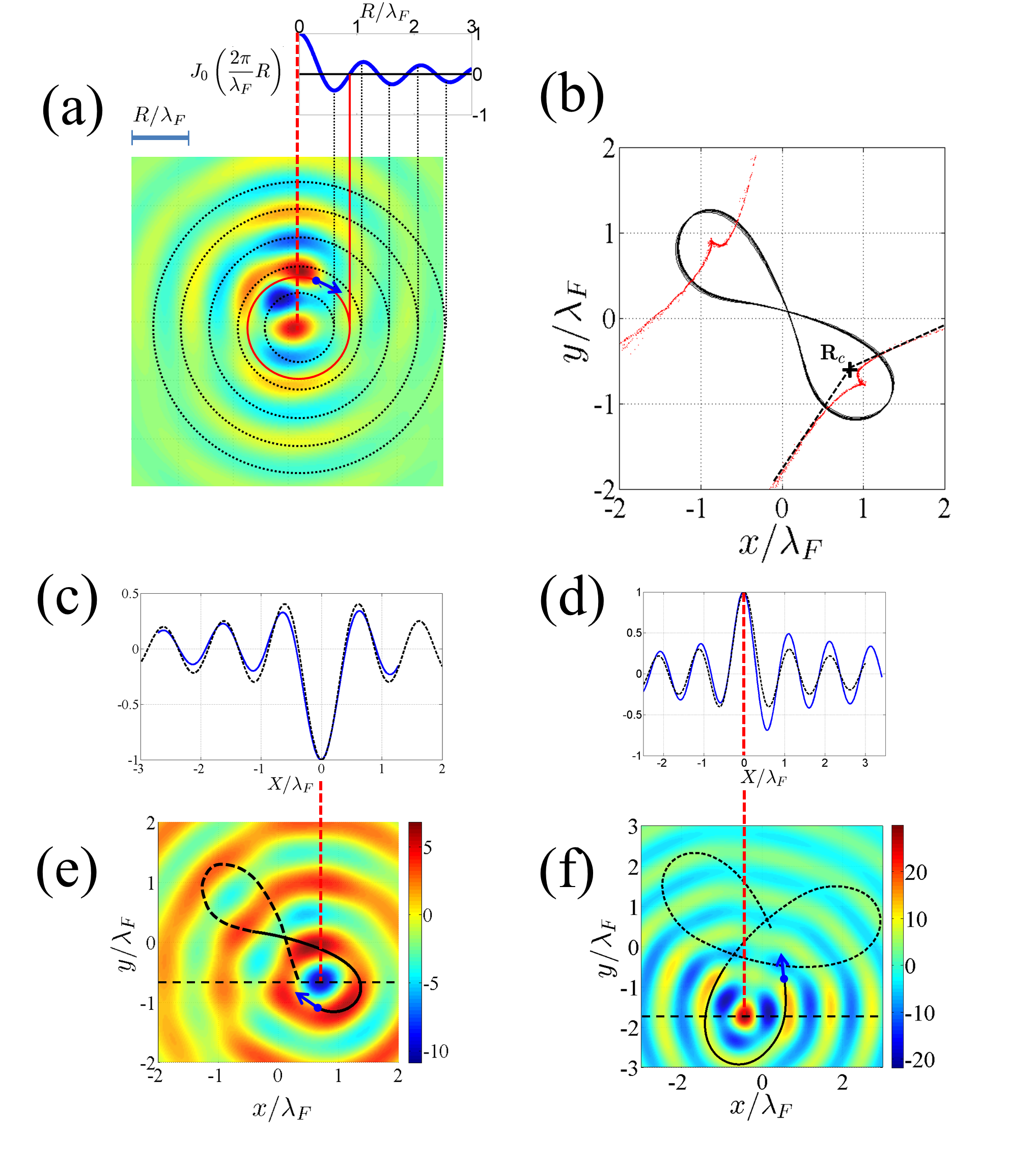}
\caption{(a) Simulated circular motion $(n,m)=(2,-2)$ at $M=21$ (\cite{Perrard_natureC_2014}, supplementary methods) for $\omega/2\pi= 0.023$ (in Faraday period time unit). The position of the drop is indicated by a blue spot and the whole circular motion by the red circle of radius $\simeq 0.9\lambda_F$. The circles in black dotted lines corresponds the extrema of the Bessel function of order $0$ centered at the origin and plotted in the above sub-figure. The surface field is reconstructed \textit{a posteriori} owing to Equation \ref{integralechamp} and is similar to the experimental one. We can separate the surface field into two terms: an intense part, immediately following and propelling the drop and a coherent part resulting from the interference of the secondary sources. (b) Simulated lemniscate motion $(n,m)=(2,0)$  at $M=21$ \cite{Perrard_natureC_2014} for $\omega/2\pi= 0.033$ (in Faraday period time unit). The black line indicates the path followed by the drop. The red points denote the loci of the instantaneous centre of curvature. $\vR_c$ denotes the centre of the pivotal field and correspond to the barycentre of the damped past sources. (c) and (e) Pivotal structure of a lemniscate. (c) In the blue line, the slice of the wave field; in the black line, a centred Bessel function of order $0$ of negative amplitude.  From $\vR_c$: the symmetry of the efficiently-contributed path is roughly rotational-invariant and generates the surface field corresponding to this symmetry at the leading order. (e) We superpose the reconstructed wave field and the path ($\omega/2\pi= 0.033$ (in Faraday period time unit)). The black line qualitatively indicates the part of the path effectively contributing to the wave field (typically a dimensionless memory curvilinear length $\simeq VMT_F/\lambda_F$). As the drop is turning back to the center, it generates a wave field a Bessel function of order $0$ centred in $\vR_c$. (d) and (f)  The pivotal structure of a trefoil $(n,m)=(4,2)$. (d) and (f) similar to (c) and (e) but for a trefoil ($\omega/2\pi= 0.016$ (in Faraday period time unit)).}
\label{pivotfig}
\end{figure*} 
Distinguishing the different time scales of the dynamics requires separating two physical effects taking place at the surface, the propulsion and the construction of an intermediate/long time coherent surface field structure. To highlight the separation of these two physical effects, the path and the surface field corresponding to the circular attractor $(n,m)=(2,2)$ are shown in figure \ref{pivotfig}(a). It was obtained from path memory simulations (see \cite{Perrard_natureC_2014} supplementary materials). As the memory and the position of each impact are known, the surface field can be numerically reconstructed from equation \ref{integralechamp}. Two distinct parts of the field can be distinguished in figure~\ref{pivotfig}(a), the first one, intense, directly following the drop and providing the propulsion, and the second one arising from the constructive interference of the secondary sources left all along the path. This coherent surface field structure has a dominant term reflecting the signature of the rotational symmetry of the path, here $J_0(2\pi R/\lambda_F)$ as emphasized by sub-figure \ref{pivotfig}(a). The propulsion is short-term acting, while the constructive interference requires a much larger duration to be established. This separation of time scales can be revisited in a theoretical manner considering the work of Oza \etal \cite{Oza_JFM_2_2013} and M. Miskin~\cite{Miskin_EPL_2014} for a circular path of radius $r_c$ of angular speed $\omega_c$. The coupling with the local slope of the surface field $-C \bnabla h^{\mathrm{circle}} $ can be theoretically derived, leading to

 \begin{equation}
\left\{
\begin{array}{ll} 
    \displaystyle\partial_T h^{\mathrm{circle}}=-\frac{1}{\omega_c r_c} \left(1-J_0^2(r_c)\right)+\mathcal{O}\left(\frac{1}{(\omega M)^2}\right)&(a)\\
    \displaystyle \partial_N h^{\mathrm{circle}} =  M J_0(r_c)J_1(r_c)+\mathcal{O}\left(\frac{1}{\omega^2 M}\right)&(b)
\end{array}
\right.
\label{gradientchampcirculaire}
\end{equation}

where $\partial_N$ and $\partial_T$ denotes the gradients along the normal and tangential direction to the trajectory.  The tangential gradient of the field (equation \ref{gradientchampcirculaire}(a)) becomes independent of the memory $M$ in the long memory limits. It indicates that the propulsion is a short time effect (further indexed $st$). On the contrary, the growth of the normal gradient (equation \ref{gradientchampcirculaire}(b)) with the memory $M$ indicates a longer time effect (further indexed $lt$). The propulsion slightly depends on the memory at the leading order as would be a short-acting effect. On the opposite, it takes at least few memory times to establish the constructive interferences leading to (Eq. \ref{gradientchampcirculaire}(b)). Guided by this idea, it becomes natural to decompose the field for any path into two parts
\begin{equation}
h=H^{\mathrm{st}}+H^{\mathrm{lt}}
\label{timescaledecompositionofthefield}
\end{equation}
and dissociate the propulsive contribution $H^{\mathrm{st}}$ from the coherent structure $H^{\mathrm{lt}}$. The short time effect, being mainly propulsive, admits mainly a tangential component and also imposes
\begin{equation}
    \begin{array}{ll} 
        \partial_N H^{\mathrm{st}}\simeq 0 ,& \mathrm{hypothesis}\;(a)
     \end{array}
\label{hypothesea}
\end{equation}
As a corollary, the propulsion being a short term action, we define the long term evolution of the field so that it does not contribute anymore to the propulsion, \textit{i.e.}
\begin{equation}
    \begin{array}{ll} 
        \partial_T H^{\mathrm{lt}}= 0 ,& \mathrm{hypothesis}\;(b)
     \end{array}
\label{hypotheseb}
\end{equation}
These two hypotheses impose two constraints which should have consequences on the description of the field. The Frenet decomposition of the surface wave (equation~\ref{decompositionchampsemilocal}) prescribes
\begin{equation}
\begin{array}{ll}
\displaystyle h\left(\vradius,t \right) &=h_0^{F} f_0^{\mathrm{F}}(t)+\sum\limits_{n\in \mathbb{Z}^*} h_n^{F} f_n^{F}(t)\\
&=h_0^{F}J_0\left(R(t)\right)+\sum\limits_{n\in \mathbb{Z}^*} h_n^{F} J_n\left(R(t)\right)e^{in\psi(t)}
\end{array}
\label{identificationterm}
\end{equation} 
Seeking in equation~\ref{identificationterm}, the terms having a zero tangential derivative ($\mathrm{hypothesis}\;(b)$) yields $H^{\mathrm{lt}}=h_0^{F} f_0^{\mathrm{F}}(t)$. Consequently, we identify $H^{\mathrm{st}}=\sum_{n\in \mathbb{Z}^*} h_n^{F} f_n^{F}(t)$. Note that the hypothesis (b) can be taken as a definition of $H^{\mathrm{lt}}$. $H^{\mathrm{st}}$ contains all the tangential derivatives and is the only one responsible of the propulsion. Section~\ref{Short} is devoted to this short time scale dynamics and will express $\partial_T H^{\mathrm{st}}$ on its symmetry in speed. But what about the hypothesis (a)? As we remain in the short memory regime, the tangential component of the gradient $\partial_T H^{\mathrm{st}}$ dominates its normal component $\partial_N H^{\mathrm{st}}$~\cite{Labousse_PRE1_2014}. But is it still valid as we get into longer memory regime? Section~\ref{intermediate} deals with the intermediate time scale dynamics and  will investigate how this physical hypothesis stands as we go onto a more complex trajectory.

\section{Short time scale dynamics: the propulsion \label{Short}}
The income of energy to the drop mediated by $H^{\mathrm{st}}$ increases its horizontal momentum. Nevertheless, as the friction starts acting, the drop loses a part of energy and a dynamical exchange of its energy between the surface field and the drop occurs. This process defines a steady velocity $v_0$ at which the energy propelling the particle balances its frictional loss of energy. In the regime parameter of interest, this equilibrium speed is a constant of the motion. In the phase space of the dynamics, this constraint defines a manifold in the neighbourhood of which it is convenient to express the dynamics. This constraint on speed contains some of the major non Hamiltonian features of the dynamics in the short memory regime as shown in \cite{Labousse_PRE1_2014}.  
Experiments and path-memory simulations of the dynamics in a harmonic potential \cite{Perrard_natureC_2014} showed that for a given drop, the mean velocity is a constant of the motion and the corresponding fluctuations remain small (typically$\sim 10\%$). In the tangential direction, the combined effect of friction and propulsion from the surface $\mathbf{f}(\vv)$ can be expressed as 
\begin{equation}
\mathbf{f}(\vv)=-\gamma \vv-C\partial_T h \vT=-\left(\gamma v+C\partial_T H^{\mathrm{st}}\right)\vT. 
\end{equation}
Let us mention that only the short term field $H^{\mathrm{st}}$ contributes in the tangential direction as $\partial_T H^{\mathrm{lt}}=0$. $\mathbf{f}(\vv)$ is tangential to the trajectory \textit{i.e.} $\mathbf{f}(\vv)=f(\vv)\vT$. The amplitude of the propulsion $f(\vv)$ must provide $v_0$ as fixed point and also is only a function of the amplitude of the speed \textit{i.e.} $\mathbf{f}(\vv)=f(v)\vT$. Additionally  $f(v)$ must be odd in $v$ as a complete reverse of the instantaneous of the motion should not break any symmetry in the propulsion, \textit{i.e.} $f(-v)=-f(v)$. It has been shown \cite{Labousse_PRE1_2014} that at the smallest order of fluctuations, the propulsion should be written as 
\begin{equation}
f(v)=\Gamma v\left(1-\left(v/v_0\right)^2+\mathcal{O}\left(v/v_0\right)^4 \right)
\label{selpropulsionterm}
\end{equation}
$\Gamma$ is a dimensionless coupling constant which can eventually depends on the memory. Equation~\ref{selpropulsionterm} known as a Rayleigh-type friction  \cite{Labousse_PRE1_2014,Rayleigh_1877,Erdmann_2005} can be understood from a dynamical point of view: if $v$ the speed of the particle decreases below $v_0$, the density of secondary sources left behind the particle increases the efficiency of the surface field propulsion. On the contrary, if $v$ increases above its set point $v_0$, the density of secondary sources decreases, and the propulsion loses some of its efficiency. This asymptotic expansion of the propulsion is similar to the propulsion model of A. Boudaoud {\it et al} \cite{Walker_Nature} and can also be derived from the theoretical works of Oza \etal \cite{Oza_JFM_1_2013}. An extension of this model considering the fluctuations in speed has been recently developed by Bush \etal   \cite{Bush_JFM_2014} and shows the importance of these higher order terms in recovering effective mass effects. The propulsion being expressed in this simple manner, how does the rest of the surface field contribute to the emergence of the intermediate time scale dynamics and of the associated pivotal wave structure? 

\section{Intermediate time scale dynamics: a semi local organization\label{intermediate}}
We study in this part, the construction of the semi-local wave structures arising at intermediate time scale. We apply this approach to the lemniscate attractor in the case of a (2D) harmonic potential. First, we propose a qualitative approach. Second we use the time scale decomposition of the field decomposition (equation~\ref{timescaledecompositionofthefield}) suggested in section~\ref{intro2} to rationalize the intermediate time scale dynamics.

\subsection{A qualitative approach}
Figure \ref{pivotfig}(b) shows the lemniscate attractor $(n,m)=(2,0)$ in a stable form obtained by path memory simulations (\cite{Perrard_natureC_2014}, supplementary methods), similar to the experimental one. We also indicate (in red points) the location of the instantaneous centre of curvature. As shown in \cite{Perrard_natureC_2014,Perrard_PRL_2014}, the elementary lemniscate path exits under other forms, with an azimuthal drift (typically smaller than $\sim 20^{\circ}$ per orbital period) or by intermittency in a chaotic regime. In all cases however and as shown in figure \ref{pivotfig}(b), when the particle turns back, its speed is minimal and its average centre of curvature concentrates in the neighbourhood of the pivotal point, noted $\vR_c$. In figure \ref{pivotfig}e, we overdraw in black curve the part of the path contributing effectively to the pivotal surface structure (typically the $M$ last secondary sources). Then, we plot in the associated figure \ref{pivotfig}c (in blue line) a slice of the normalized surface field along the $x$-axis, and in dashed line a Bessel function of order zero centred at the pivotal point $\vR_c$ and of negative and normalized amplitude. The field generated by this elementary pivotal motion is a pivotal field and is well fitted by a Bessel function of order $0$, centred in $\vR_c$ which actually defines $\vR_c$. Let us  note that the emergence of this structure does not require a large number of secondary sources. The same phenomenon occurs for the trefoil attractor $(n,m)=(4,2)$ as indicated in the coupled Figures \ref{pivotfig}(d) and (f).\\

Locally the turn back generates a surface field $\pm J_0(\Vert \vradius-\vR_c \Vert)$: the semi local symmetry of the surface field is a signature of the semi local  circular symmetry of the path. Qualitatively, the effect of this process is the emergence of a preferred radius of curvature. We now derive theoretically how such structures arise.

\subsection{A theoretical approach}
The short time scale provides information in the tangential direction and it has been shown in the paragraph \ref{Short} that the propulsion is mediated by short time field $H^{\mathrm{st}}$. The rest of field $H^{\mathrm{lt}}$, revealed as the memory increases, is mainly involved in the normal mechanical balance. We now investigate its role in the emergence of the intermediate time scale dynamics and in the construction of the pivotal wave structure.\\
   
$(\vT,\vN)$ denotes the direct Frenet basis (see figure \ref{schema}(a)). We note $\mathcal{T}=\vradius . \vT$ and $\mathcal{N}=\vradius . \vN$, evolving in time by definition as $\dot{\mathcal{T}}=v+\mathcal{N} v/R$ and $\dot{\mathcal{N}}=-\mathcal{T} v/R$. The mechanical tangential balance yields $\dot{v}=  \Gamma v\left(1-v^2/v^2_0 \right)-\partial_T E_p/m$ , where $\partial_T E_p$ denotes the tangential projection of the external potential. The normal mechanical balance involves the normal gradient of the field. As $\partial_N H^{\mathrm{st}}=0 $, the normal mechanical balances can be simply written as $v^2/R=-\partial_N E_p/m -C\partial_N H^{\mathrm{lt}}$.\\

The time decomposition of the surface field indicates that $H^{\mathrm{lt}}$ predominantly  contributes in the normal direction (Hypothesis (b)). It can be expressed as 
\begin{equation}
\partial_N H \simeq \partial_N H^{\mathrm{lt}}=h_0^F(t)J_1\left(R(t)\right)
\end{equation}
where $h_0^F$ is the amplitude of the mode $n=0$. Note that neglecting $\partial_N H^{\mathrm{lt}}\gg \partial_N H^{\mathrm{st}}$ is equivalent to the hypothesis (b). This hypothesis relies on the symmetry of the paths we intend to describe: at an intermediate time scale dynamics, the trajectory of the walker is made of a succession of loops. Each loop promote a dominant local symmetry. 

Also the separation of the time scales in the dynamics enables the modelling of the dynamics by a set of equations

\begin{equation}
\left\{
    \begin{array}{ll} 
       \dot{\mathcal{T}}=v(1+\mathcal{N}/R)  & (a)\\
       \dot{\mathcal{N}}=-v \mathcal{T} /R   &(b)\\
       \dot{h}_0^F=-h_0^F/M+h_0J_0\left(R \right)   & (c)  \\
       \dot{v}=  \Gamma v\left(1-v^2/v^2_0 \right)-\partial_T E_p/m & (d)  \\   
       v^2/R+\partial_N E_p/m+C J_1\left(R \right)h_0^F=0 & (e)
     \end{array}
\right.
\label{equationnormalgeneral}
\end{equation}
Equations \ref{equationnormalgeneral}(a) and \ref{equationnormalgeneral}(b) are kinematic and are consequently always valid. Equation \ref{equationnormalgeneral}(c) is a differential equation in $h_0^F$ and simply corresponds to a derivation of its integral form (equation \ref{decompositionchampsemilocal}). It means that the mode $0$ of the Frenet-adapted wave basis evolves in time owing to two distinct contributions: it relaxes over a memory time and the mode is maintained by the coupling with the drop impacts. Equation \ref{equationnormalgeneral}(c) can be taken as a definition of $h_0^F$ and thus does have be checked. Equation \ref{equationnormalgeneral}(d) is the momentum balance in the tangential direction and is a direct consequence of the expansion of the dynamics in the neighbourhood of the manifold $\left\lbrace v=v_0 \right\rbrace $. Equation \ref{equationnormalgeneral}(e) arises from the normal mechanical balance and differs from the others as no time-derivative is involved. This implicit equation in $R$ actually defines the manifold on which the dynamics evolves.\\

We apply this approach in the case of a (2D) harmonic potential $E_p=m\omega^2 r^2/2$ of eigenpulsation $\omega$ which leads to
\begin{equation}
\left\{
    \begin{array}{ll} 
       \dot{\mathcal{T}}=v(1+\mathcal{N}/R)  & (a)\\
       \dot{\mathcal{N}}=-v \mathcal{T} /R   &(b)\\
       \dot{h}_0^F=-h_0^F/M+h_0J_0\left(R \right)   & (c)  \\
       \dot{v}=  \Gamma v\left(1-v^2/v^2_0 \right)-\omega^2 \mathcal{T}  & (d)  \\   
       v^2/R+\omega^2\mathcal{N}+C J_1\left(R \right)h_0^F=0 & (e)
     \end{array}
\right.
\label{equationnormal}
\end{equation}
Also, the complexity of the dynamics has been reduced to a five dimensional representation $(\mathcal{T} ,\mathcal{N},v,h_0^F,R)$. Can we trust the set of equations~\ref{equationnormal}(d) and ~\ref{equationnormal}(e)? This raises actually two distinct questions. i) Is the equation~\ref{equationnormal}(d) model correctly the propulsion? It is interested to remark that equation~\ref{equationnormal}(d) relies only on speed symmetry arguments which should remain valid for various memory regimes. The question has been already successfully addressed in the low memory regime by Labousse \textit{et} Perrard~\cite{Labousse_PRE1_2014}, but up till now, remains unproven at larger memory. It is the first point we have to check. ii) The simplicity of equation~\ref{equationnormal}(e) actually relies on hypothesis (a) (Eq.~\ref{hypothesea}). It is natural to retain only the dominant field symmetry corresponding to a single pivotal motion but we have to evaluate \textit{a posteriori} the relevance of such an hypothesis. 

\begin{figure*}
    \centering
\includegraphics[width=\columnwidth]{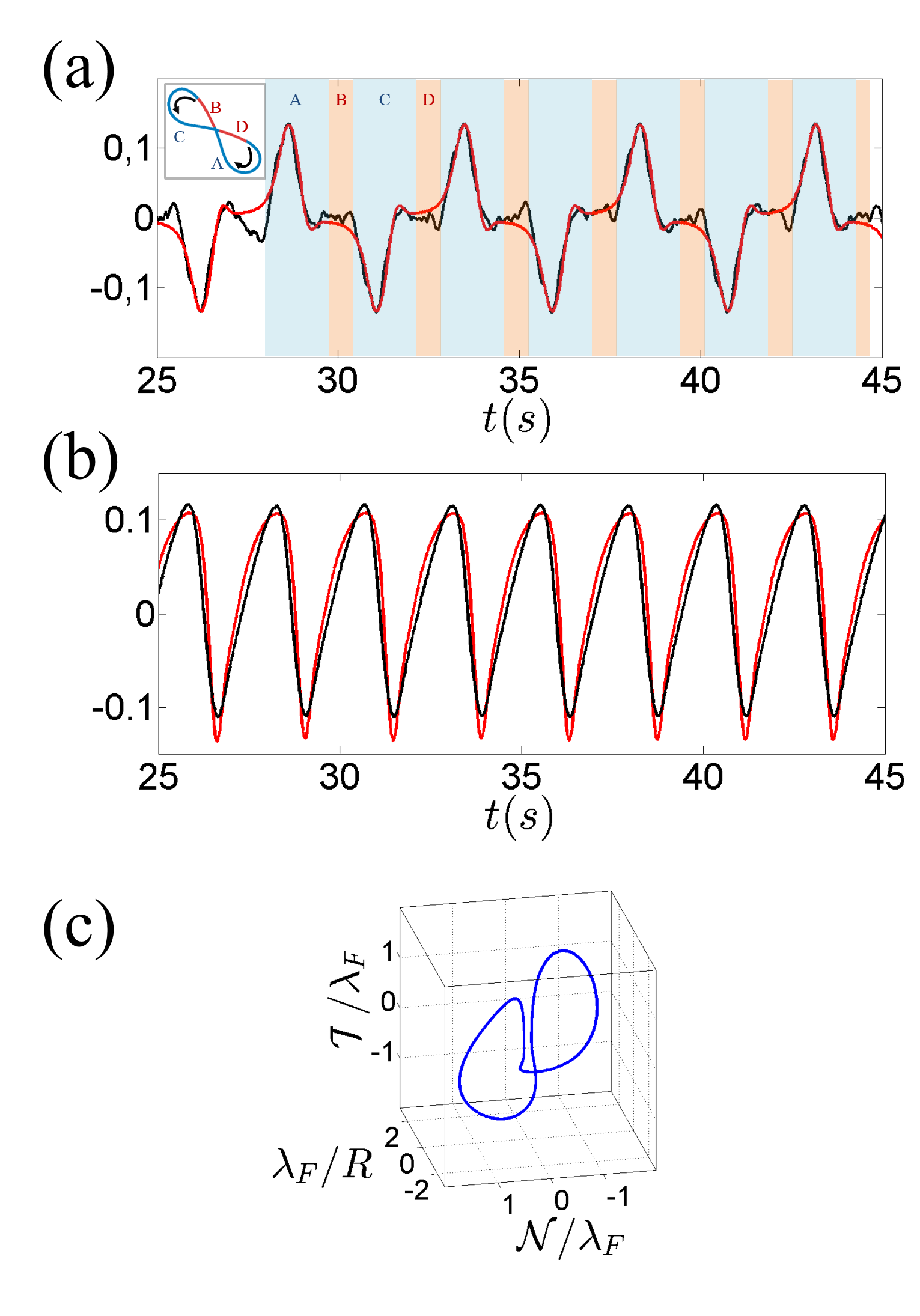}
\caption{Verification of equation \ref{equationnormal} for a lemniscate extracted from the Fort numerical model ($\omega/2\pi= 0.033$ (in Faraday period time unit), $M=21$). The time and length scales are expressed in natural units. (a) Normal balance from simulation results: in the black line $-CJ_1(2\pi R/\lambda_F)h_0^F$ with $C=1.40$ m.s$^{-2}$ ($C=0.18$ in dimensionless units), in the red line $v^2/R+\omega^2\mathcal{N}$. The blue time intervals (A,C) indicates a good theoretical predictions of equation~\ref{equationnormal}(e) and correspond to the build up of a pivotal field. The red time intervals (B,D) correspond to the parts of the path that do not adequately fits with equation~\ref{equationnormal}(e).  The inset is the simulated path (see figure~\ref{pivotfig}(b)) and links the time intervals (A,B,C,D) to their corresponding portion of path.  (b) Tangential balance from simulation results: in the black line $\dot{v}+\omega^2\mathcal{T}$, in the red line $\Gamma v(1-v^2/v_0^2)$ with $\Gamma=35.0$  s$^{-1}$ ($\Gamma=0.875$ in dimensionless units) and $v_0=\sqrt{<v^2>}\simeq=7.9$ mm.s$^{-1}$ (or $v_0=4.2\times 10^{-2}$ in dimensionless units). Fluctuations of speed are of $17\%$. (c) Projection of the attractor in the dimensionless representation $(\lambda_F/R,\mathcal{N}/\lambda_F,\mathcal{T}/\lambda_F)$}
\label{verifsimu}
\end{figure*}

The relevance of the set of equation \ref{equationnormal} is checked for a lemniscate trajectory in figure \ref{verifsimu}. Figures \ref{verifsimu}(a) and \ref{verifsimu}(b) respectively, represent the time evolution of the terms involved in the normal and tangential directions, (respectively equations~\ref{equationnormal}(e) and~(d)). The normal balance is well captured by the equation~\ref{equationnormal}(e)  (blue time intervals of Fig.~\ref{verifsimu}a) and the neglected terms $\sim \partial_N H^{\mathrm{st}}$ are well negligible. This justifies hypothesis (a) (see equation~\ref{hypothesea}). As expected, we also note that the Frenet basis decomposition is not convenient as the walker moves in a quasi straight line motion (red time intervals of figure~\ref{verifsimu}(a)). Figure~\ref{verifsimu}(b) shows that the tangential balance is also well captured by the equation~\ref{equationnormal}(d). This paragraph also justifies the physical distinction between a short time field $H^{\mathrm{st}}$ mainly involved in the propulsion, and a long time field $H^{\mathrm{lt}}$ mainly involved in the normal balance. It explains the relevance of the decoupling of the effects of the two distinct time scales as it can capture the main features of the dynamics. 

Figure \ref{verifsimu}c represents a three dimensional projection of lemniscate attractor $(n,m)=(2,0)$. The constraint on speed enables a large reduction of dimensions of the phase space in comparison to the integro-differential formulation in \cite{Molacek_JFM_2_2013,Oza_JFM_1_2013}. In the integro-differential formulation, the wave field stores an infinite number of degree of freedom. Provided the fluctuations of speed remain small, a large number of these dimensions can be reduced in a resulting propulsion force (equation \ref{selpropulsionterm}) for a large range of memory parameter.\\
It is useful to study some particular cases of equation \ref{equationnormal} to further analyse the consequences of the speed constraint. A first interesting particular case  is the simplest fixed point $v^*=v_0$. It induces $\mathcal{T}^*=0$ from \ref{equationnormal}(d) then $\dot{\mathcal{N}}^*=0$ from \ref{equationnormal}(b), $\mathcal{N}^*=-R^*$ from \ref{equationnormal}(a) and finally $h_0^{F*}=MJ_0(R^*)$, that corresponds to a circular attractor. The constraint 
\begin{equation}
((v_0/R^*)^{2}-(\omega)^2)R^*+CM J_1\left(R^* \right)J_0\left(R^* \right)=0
\end{equation}
is multi-valued as several $R^*$ can satisfy this equation in the high memory limit. The term $\delta=(v_0/R^*)^{2}-(\omega^*)^2$ corresponds to a mismatch of frequency between the one arising from the external harmonic potential $\omega$ and the natural frequency $v/R$ prescribed by the surface field. They are not necessary equal which is a signature of the difference of symmetry between the surface field force and the external force: this set of equations could provide a simplified theoretical framework to study the transitions to the low-dimensional chaotic regimes reported by Perrard \textit{et al.}~\cite{Perrard_PRL_2014}. We recover the condition $J_0(R^*)J_1(R^*)=0$ in the long memory limit which gives rise to a quantization of the radius of curvature \cite{Fort_PNAS,Perrard_natureC_2014,Miskin_EPL_2014,Oza_JFM_2_2013,Harris_JFM_2014,Oza_PoF_2014}.\\

Another interesting limit case arises when there exists a strong mismatch between the centre of curvature and the origin of the external force $\mathcal{N}/R \ll 1$. Equation \ref{equationnormal}(d) shows that any spread of the speed from its set point induces a retroaction from the term $v(1-v^2/v_0^2)$. Nevertheless the coupling to the external potential may induce a non-stationary solution in speed. Indeed, provided $\mathcal{N}/R \ll 1$, time-derivating equation \ref{equationnormal}(d) leads to $\ddot{v}+\Gamma\dot{v}(1-3v^2/v_0^2)+\omega^2v=0$ and provides a self oscillation of speed (Van der Pol type). These fluctuations of speed would make a straight line motion unstable to any transverse fluctuations. This instability of the straight line motion is observed in harmonic potential~\cite{Perrard_natureC_2014,Labousse_PRE1_2014} in the reported parameter regimes.\\

This paragraph focused on the intermediate time scale dynamics. We adopted a point of view adapted to the symmetry of the wave field by choosing to develop the surface field into a Frenet-adapted wave basis. It enables us to separate a short time effect, the propulsion, and an intermediate time effect, the tendency of the dynamics to build pivotal field and the emergence of preferred radii of curvature. What will happen to these pivotal structures at the long time scale dynamics? 

\section{Long time scale dynamics: The global organization of the pivots\label{Long}}
We zoom out once again in the scales of time and observe the dynamics at a longer time scale. The memory is thus long enough to store several pivotal structures in the wave field. These pivotal structures should interact via the walker's trajectory. One would also expect to feel the effect of the central force and of its relating symmetry. Excepted the circular attractor, how can the construction of pivotal field be compatible with the axisymmetry of the (2D) central harmonic potential?\\
 
\begin{figure*}
    \centering
\includegraphics[width=\columnwidth]{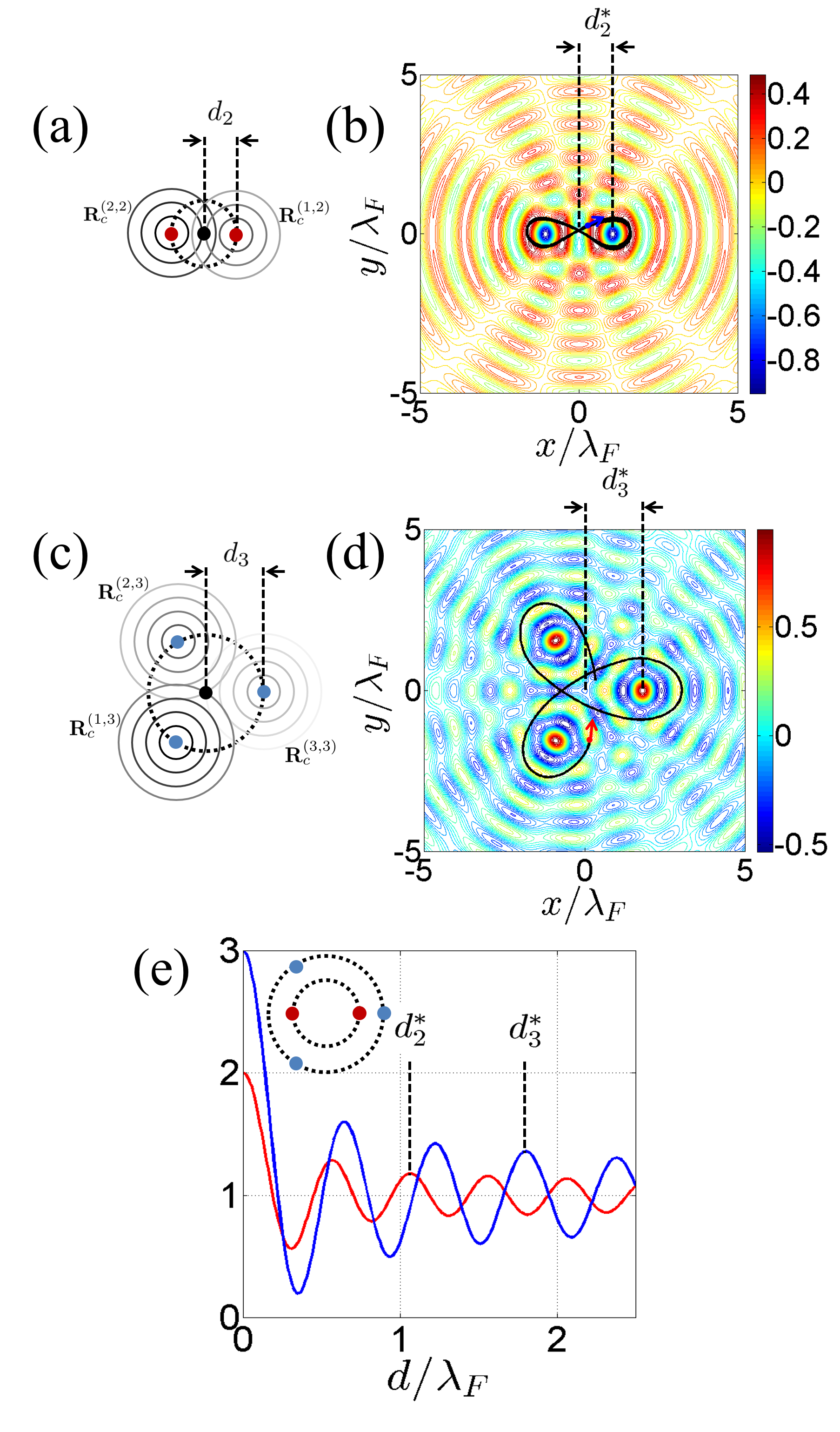}
\caption{(a) Sketch of the interaction between two pivotal fields. The pivotal points (two red dots) are diametrically opposed on a circle of radius $d_2$. The black dot indicates the centre of the harmonic potential. (b) Superposition of two pivotal fields $\mathcal{H}_2=-(J_0(\Vert \vradius -\vR_c^{(1,2)} \Vert) +  J_0(\Vert \vradius -\vR_c^{(2,2)} \Vert))$ and the associated path with $d_2=\Vert\vR_c^{(1,2)}\Vert=\Vert\vR_c^{(2,2)} \Vert $. Here, is chosen a field with $d_2=d_2^*$. The lemniscate path is identical to the figure~\ref{pivotfig}(e). (c) Sketch of the interaction between three pivotal fields. The pivotal points (three blue dots) are $2\pi/3$ equally spaced on a circle of radius $d_3$. The black dot indicates the center of the harmonic potential. (d) Superposition of three pivotal fields $\mathcal{H}_3=J_0(\Vert \vradius -\vR_c^{(1,3)} \Vert) + J_0(\Vert \vradius -\vR_c^{(2,3)} \Vert)+ J_0(\Vert \vradius -\vR_c^{(3,3)} \Vert)$ and the associated path with $d_3=\Vert\vR_c^{(1,3)}\Vert=\Vert\vR_c^{(2,3)}\Vert=\Vert\vR_c^{(3,3)}\Vert$. Here, is chosen a field with $d_3=d_3^*$. The trefoil path is identical to the figure~\ref{pivotfig}(f). (e) Evolution of $I_2(d_2)$ (red line) and $I_3(d_3)$ (blue line) with $d$. Here, $d$ is common notation for $d_2$ or $d_3$}
\label{pivotexpsimu}
\end{figure*}

Perrard \textit{et al.}~\cite{Perrard_natureC_2014} show that the lemniscates and the trefoils attractors have quantized extensions. Consequently, the location of these pivotal points should be well defined, meaning that they adopt a particular position in space. We propose a simple geometrical approach to justify this quantization. 
Each about-turn generates a Bessel function centred at its associated pivotal wave structure $J_0(\Vert \vradius -\vR_c^{(k,n)}\Vert )$. $n$ indicates the number of pivotal point: $n=2$ for the lemniscate, $n=3$ for the trefoil and $k=1,\ldots,n$ denote the $k$-th pivotal points.
We sketched on Fig.~\ref{pivotexpsimu}a, the position of the pivotal points $(\vR_c^{i,2})_{i=1,2}$ of a lemniscate. They are diametrically opposed on a circle of radius $d_2$. We superpose on figure~\ref{pivotexpsimu}(b) a numerical lemniscate path to a field made of the superposition of two pivotal fields, each of them centered on a pivotal point $(\vR_c^{i,2})_{i=1,2}$.  Figures \ref{pivotexpsimu}(c) and (d) are similar to figures \ref{pivotexpsimu}(a) and (b) except for a trefoil, with three pivotal points $(\vR_c^{i,3})_{i=1,2,3}$ that are $2\pi/3$ equally spaced.

Figure \ref{pivotexpsimu}(b) indicates  the contour line of the superposition of two pivotal fields $\mathcal{H}_2=-(J_0(\Vert \vradius -\vR_c^{(1,2)} \Vert) + J_0(\Vert \vradius -\vR_c^{(2,2)} \Vert))$ and figure \ref{pivotexpsimu}(d) shows the superposition of three pivotal fields $\mathcal{H}_3=J_0(\Vert \vradius -\vR_c^{(1,3)} \Vert) + J_0(\Vert \vradius -\vR_c^{(2,3)} \Vert)+ J_0(\Vert \vradius -\vR_c^{(3,3)} \Vert)$. To estimate the spatial coherence between the successive pivotal fields, we compute the interfering quantity related to the superposition of $n$ pivotal fields.
\begin{equation}
\begin{array}{ll}
\displaystyle I_n(d_n)&= \displaystyle\frac{1}{A_n}\int\int \mathcal{H}_n^2 dS\\
&\displaystyle = \frac{1}{A_n}\int\int \left(\sum\limits_{k=1}^nJ_0(\Vert \vradius -\vR_c^{(k,n)} \Vert) \right)^2  dS.
\end{array}
\end{equation}
where $d_n=\Vert \vR_c^{(k,n)} \Vert $ is the distance between the $k$-th pivotal point and the center of symmetry of the trajectories. The integration is realized numerically in a large $(20\lambda_F\times 20\lambda_F)$ but finite centred surface domain. It is normalized by $A_n$ a quantity independent of $d_n$ which makes $I_n$ independent on the domain of integration. 
 \begin{equation}
A_n=\frac{1}{n}\int\int \left(\sum\limits_{k=1}^nJ_0(\Vert \vradius \Vert) \right)^2  dS.
\end{equation}
Figure \ref{pivotexpsimu}(e) shows the evolution of $I_n$ with $d_n$ and presents optima of interfering conditions for quantized distances $d_n=d^*_n$. In particular, it predicts $d^*_2=1.08\lambda_F$ for the lemniscate and $d^*_3=1.80 \lambda_F$ for the trefoil, respectively. It is in good agreement with the pivotal points of the lemniscate in figure \ref{pivotfig}(e) ($1.015\lambda_F$), and of the trefoil in figure \ref{pivotfig}(f) ($1.79 \lambda_F$). The well-defined position of the pivotal points are a signature of the symmetry of the external potential. It suggests reconstructing the equation of motion in a basis adapted to the harmonic central potential. It would imply that $\mathcal{H}_n(d^*)$ corresponds to a local maximum of overlapping with the central wave basis $\left\lbrace f_n^{\mathrm{ext}}(t)\right\rbrace_{n\in \mathbb{Z}} =\left\lbrace J_n(r(t)e^{in\theta(t)})\right\rbrace _{n\in \mathbb{Z}}$. Figure \ref{pivotexpsimu}(f) presents other maxima but they are not observed experimentally. For low distance $d_n$ these states are in competition with the circular attractors. This geometrical approach only demonstrates how the pivotal fields are organized between each other but did not account for the dynamical stability. 
  
\section{Conclusion}
We investigated the walker dynamics subjected to an attractive potential. The main goal of the current paper was to show that the dynamics lies on three time scales unveiled as the memory parameter increases. At short time scale, the accompanying surface wave simply propels the drop. At an intermediate time scale, the appearance of coherent wave structure that we call pivotal field, induces motion with preferred radii of curvature. This elementary wave pattern gives rise to well-defined motion and can be seen as an elementary block of motion. The long time scale dynamics assembles these elementary pivotal fields to maximize their mutual interferences. The external potential imposes its own symmetry which a given disposition of the pivotal fields has to account for.\\

We applied our theoretical framework to the particular case of the (2D) harmonic potential. We explained how these three time scales and relating spatial self-organizations are interlocked. The mechanism provides a new rule of construction for the macroscopic eigenstates reported in \cite{Perrard_natureC_2014}.\\

The research of coherent structures is a common technique in complex systems, like in fully developed turbulent flows. In this latter case, the details of the dynamics does not really matter and much of the information is stored in a much simpler structure that leads to a hierarchy in the description of the complexity. In the current article, the details of the reported dynamics are forgotten as they contribute to the emergence of higher order structures. One can reason from a "dynamics" point of view, but will be rapidly limited as the complexity of description increases tremendously. One can also sacrifice some details of the dynamics and adopt a description at another time scale. In this limit, the problem is reduced to a self-organization of the elementary wave patterns. The useful amount of information (said differently the apparent dimensions of the system), depends here on the time scale of the description. In the current object of study, it explains why the dynamics, apparently complex and chaotic, can be reduced to its attractors that can be simply defined by two integers and called eigenstates. 

\acknowledgments{The authors thank Marc Miskin, John Bush, Ruben Rosales and Anand Oza for useful discussions. This research is supported by the AXA Research Fund and the french Agence Nationale de la Recherche, through the project "ANR Freeflow", LABEX WIFI (Laboratory of Excellence ANR-10-LABX-24) within the French Program “Investments for the Future” under reference ANR-10- IDEX-0001-02 PSL.
}

\end{document}